# Skyrmions in the moiré of van der Waals 2D magnets

Qingjun Tong[†], Fei Liu[†], Jiang Xiao[‡], and Wang Yao[†,*]

[†]Department of Physics and Center of Theoretical and Computational Physics, University of Hong Kong, Hong Kong, China

[‡]Department of Physics and State Key Laboratory of Surface Physics, Fudan University, Shanghai 200433, China

*Correspondence to: wangyao@hku.hk

**Abstract:**

We explore the skyrmion formation and control possibilities in 2D magnets from the ubiquitous moiré pattern in vdW heterostructures. Using the example of a ferromagnetic monolayer on an antiferromagnetic substrate, we demonstrate a new origin of skyrmions in the 2D magnets, from the lateral modulation of interlayer magnetic coupling by the locally different atomic registries in moiré. The moiré skyrmions are doubly degenerate with opposite topological charge, and trapped at an ordered array of sites with the moiré periodicity that can be dramatically tuned by strain and interlayer translation. At relatively strong interlayer coupling, the ground states are skyrmion lattices, where magnetic field can switch the skyrmion vorticity and location in the moiré. At weak interlayer coupling limit, we find metastable skyrmion excitations on the ferromagnetic ground state that can be deterministically moved between the ordered moiré trapping sites by current pulses. Our results point to potential uses of moiré skyrmions both as information carriers and as drastically tunable topological background of electron transport.

Recent discovery of monolayer ferromagnets has provided unprecedented opportunities for exploring magnetisms and their spintronic applications in the two-dimensional (2D) limit[1,2]. Atomically thin magnetic tunnel junction devices[3-7], and gate-tunable magnetism[8-11] are already demonstrated in van der Waals (vdW) heterostructures made with the 2D magnets. In vdW heterostructures, the way functional layers stack is critical in determining the materials properties. For example, long-period moiré pattern with the spatially varying interlayer atomic registry can lead to the observations of various intriguing topological and correlated phenomena in graphene[12-18], and lateral patterning of electronic structures in transition metal dichalcogenides[19]. For $CrI_3$ where monolayer is discovered as an Ising ferromagnet[2], first principle studies suggest that a homobilayer can have either antiferromagnetic (AFM) or ferromagnetic (FM) interlayer order depending on the interlayer registry[5,20-22]. The registry dependent magnetic coupling is also expected in the vdW heterostructure formed between a FM monolayer and a magnetic substrate, especially when the latter has a lateral AFM order[23-28]. In the context of moiré pattern formation, the atomic registry dependence in the interlayer magnetic coupling and ordering points to the intriguing possibility of magnetization textures defined by moiré.

Skyrmion is a vortex or anti-vortex like magnetization texture protected by nontrivial topology which has attracted intense interest for their promising potentials in spintronics technology[29-36]. In most systems observed[32-35], skyrmions originate from the chiral Dzyaloshinskii-Moriya interaction (DMI) that exists only under broken inversion symmetry and strong spin-orbital coupling[29-31], which has limited the material choices. New and general mechanisms for skyrmion formation and new means of skyrmion control are highly desired for its exploration in the emerging arena of vdW 2D magnets.

Here we explore the skyrmion formation and control possibilities from the long-period moiré pattern in vdW heterostructure of 2D magnets. The skyrmion formation can be rooted in the local-to-local variation of atomic registry in the moiré and the registry dependent interlayer magnetic coupling, which we demonstrate using the example of a FM monolayer on a layered AFM substrate with lateral Néel order. We show that, under the moiré modulation, interlayer exchange coupling favors periodic domain structures in the monolayer magnet, whereas interlayer dipolar coupling can wind domain boundaries to form skyrmions. At large moiré periodicity, the ground states are doubly degenerate skyrmion lattices characterized by opposite topological charge. An external magnetic field can switch the vorticity and location of skyrmions in the moiré. The skyrmion lattices can be dramatically tuned by small interlayer translation and strain through the moiré magnification effect. These form the basis of magnetic and mechanical control of the moiré skyrmion lattices. At weaker interlayer coupling or small moiré period where the monolayer remains in FM phase, we find that local magnetization reversal can generate and erase metastable skyrmion excitations that behave like quasiparticles trapped in a superlattice potential. Current pulses can deterministically move such skyrmion excitations between the ordered trapping sites. These control possibilities point to potential use of 2D magnets moiré as a lateral register array using skyrmion excitations as information carriers.

Fig. 1a schematically illustrates a FM monolayer on a layered AFM substrate, both of a hexagonal lattice. Lattice mismatch and/or a relative twisting between the layers lead to the formation of moiré pattern. Under small lattice mismatch $\delta$ and/or the twisting angle $\theta$, the moiré periodicity $A \approx a/\sqrt{\delta^2 + \theta^2}$ with $a$ being the lattice constant of the monolayer. In a long-period moiré, as shown in Fig. 1b, the atomic registries in different local regions, at a length scale small compared to $A$ but large compared to $a$, resemble the registries of lattice-matched stacking of different interlayer translation. A lattice-matched stacking can be characterized by an interlayer translation vector **r**, defined as the lateral displacement between two sites from the two layers (Fig. 1b inset). So the moiré in general can be described by a mapping function **r(R)** that gives the local atomic registry as a function of location **R** in the moiré.

We consider AFM substrate of lateral Néel order and perpendicular anisotropy[24-26,37]. The interlayer magnetic coupling from such substrate and the resultant magnetic ordering of the FM monolayer are highly sensitive to the atomic registry. As Fig. 1b schematically shows, the interlayer exchange coupling aligns the monolayer magnetization in opposite out-of-plane directions when the atomic sites of FM monolayer sit on top of the two AFM sublattices with opposite magnetic orders. In the moiré, the local-to-local variation in interlayer registry therefore realizes a lateral modulation in interlayer magnetic coupling that tends to generate magnetization textures in the monolayer magnet (e.g. opposite magnetizations at the two locals $R_1$ and $R_2$). This can be general in long-period moiré between different magnets of either in-plane or out-of-plane anisotropy (more cases discussed in Supplementary Notes).

The interlayer magnetic coupling has the exchange and dipolar contributions, both of which are functions of position **R** through their dependence on the local registry **r(R)**. Fig. 1c and 1d show the interlayer exchange field $\mathbf{B}_{ex}(\mathbf{R})$ and dipolar field $\mathbf{B}_{dd}(\mathbf{R})$ respectively from the AFM substrate, plotted as a function of **R** within a moiré supercell (c.f. Supplementary Note 1 for details of the calculation). The exchange coupling is isotropic so $\mathbf{B}_{ex}$ only has the component (*z*) along the direction of AFM order. The dipolar field $\mathbf{B}_{dd}$ is typically much weaker; nevertheless it also plays a key role in determining the form of magnetization texture.

The magnetization of the FM monolayer orders according to the competition between $\mathbf{B}_{ex}$ and $\mathbf{B}_{dd}$ that favor magnetization textures, and the intralayer exchange interaction that favors uniform order, which can be modeled with the effective Hamiltonian,

$$\mathrm{H} = -I\sum_{<i,j>}\mathbf{m}_i\mathbf{m}_j - K\sum_i(\mathbf{m}_{z,i})^2 - \sum_i[\mathbf{B}_{ex}(\mathbf{R}_i) + \mathbf{B}_{dd}(\mathbf{R}_i) + \mathbf{B}_{ext}]\cdot\mathbf{m}_i. \quad (1)$$

Here $\mathbf{m}_i$ is the normalized local magnetization at i-th site and $<i,j>$ runs over all nearest neighboring sites of the hexagonal lattice. $I > 0$ is the ferromagnetic intralayer exchange, and $\mathbf{B}_{ext}$ is an external magnetic field. The results presented in the main text below use $K = 0.014I$, a weak perpendicular magnetic anisotropy that corresponds to the ferromagnetic monolayer CrBr$_3$ being actively explored[6]. The relative strength between the interlayer exchange field $\mathbf{B}_{ex}$, interlayer dipolar field $\mathbf{B}_{dd}$, and intralayer exchange coupling *I*, vary between different heterostructure choices, to be determined in future experiments. So their ratios are accounted here as variables in the calculations. The steady state magnetic configurations and the dynamics are solved from the coupled Landau-Lifshitz-Gilbert equations:

$$\frac{d\mathbf{m}_i}{dt} = -\gamma\mathbf{m}_i \times \mathbf{H}_i^{eff} + \alpha\mathbf{m}_i \times \frac{d\mathbf{m}_i}{dt} \quad (2)$$

where $\mathbf{H}_i^{eff} = -\partial\mathrm{H}/\partial\mathbf{m}_i$, γ and α are the gyromagnetic ratio and Gilbert damping coefficient respectively. The ground states and excited states magnetic configurations are obtained by relaxing from various trial configurations, where we use α = 1 for faster convergence.

*Skyrmion lattices.* --- Figure 2a shows the phase diagram as a function of the moiré periodicity $A$ and strength of the interlayer magnetic coupling, at zero external magnetic field. When the interlayer coupling or the moiré period gets large, magnetic domains form at the moiré periodicity due to the dominance of the interlayer exchange field that has a sign change in a moiré supercell. Furthermore, the competition between the interlayer dipolar field and intralayer exchange at the domain walls leads to two distinct phases of magnetization textures. At the small $A$ side, the magnetic moments at domain walls are aligned in-plane by the intralayer exchange, forming a magnetic bubble lattice. At the large $A$ side where the magnetic order can change smooth enough along the domain walls, even a weak interlayer dipolar field can wind the moments to form the topologically nontrivial skyrmion lattice (see also Supplementary Note 2). The topology can be characterized through an emergent electromagnetic field $\Omega(\mathbf{r}) = \frac{1}{2}\mathbf{m} \cdot (\frac{\partial \mathbf{m}}{\partial x} \times \frac{\partial \mathbf{m}}{\partial y})$ from the magnetization texture, whose integral over a supercell defines a topological charge $C = \frac{1}{2\pi} \iint \Omega(\mathbf{r}) d^2\mathbf{r}$.[29]

We find four stable skyrmion lattice configurations with distinct combinations of their topological charge ($C = 1$ or $-1$) and vorticity ($l = 1$ or $-1$). Fig. 2c shows the energy of the skyrmion lattice configurations as a function of external magnetic field. Over a moderate range of field, the vortex-like ($l = 1$) skyrmion lattices are ground states, which become doubly degenerate (with $C = 1$ and $-1$) at $B_{ext} = 0$. This double degeneracy is the consequence of the inversion symmetries of the $\mathbf{B}_{ex}$ and $\mathbf{B}_{dd}$ components about the supercell center $\mathbf{R}_0$.

The energy level diagram in Fig. 2c features several crossing points between FM states and skyrmion lattice states, and between different skyrmion lattice states. At all these crossing points, the magnetic configurations are protected from mixing with each other by their distinct topology and/or vorticity, which ensures their adiabatic evolutions through such points even in a fast sweeping magnetic field (c.f. Supplementary Note 6).

The magnetic field can tune the size of the skyrmions, as a positive (negative) field tends to increase the area of the red (blue) domains. A skyrmion configuration becomes unstable when its boundary exceeds the moiré supercell boundary (see Fig. 2c and inset), where the domain wall will reconfigure together with those at neighboring supercells. Remarkably, the reconfiguration only happens between the vortex-like ($l = 1$) and antivortex-like ($l = -1$) skyrmion lattices with the same topological charge C. Green (orange) arrows in Fig. 2c indicate the evolution of magnetic configurations by sweeping the magnetic field up (down), for an initial state of $C = 1$. The topological charge is always conserved, while the vorticity and location (**R₁** vs. **R₂** in the moiré supercell) of skyrmions switch. Such switching is accompanied by the sign change of net magnetization $\langle m_z \rangle$, and can therefore be detected as a magnetic hysteresis in sweeping magnetic field (see Supplementary Note 6).

The skyrmion lattice ground states and metastable excited states can be exploited for information storage using both the binary topological charge C and vorticity $l$. The product $l \times C$ corresponds to the sign of the net magnetization that can be directly measured using, for example, magneto-optical effects[1,2]. When the magnetization is coupled to conducting electrons, the emergent electromagnetic field $\Omega$ generates a topological Hall current[29], where the current direction reads out C. Writing of both C and $l$ values is made possible in magnetic field. Sweeping a magnetic field can flip $l$ without changing C. Preparation of $C = 1$ ($C = -1$) skrymion lattices can be realized by first polarizing the magnetization in the down (up) direction in a strong external magnetic field and then ramping down the field for the magnetization to relax to ground state (see Supplementary Note 4).

The moiré defined magnetization textures also feature remarkable mechanical controllability. The moiré pattern magnifies a small translation between the layers into a large shift of the superlattice by an amplification factor of $A/a$. This allows dramatic tuning of the skyrmion locations through tiny interlayer translation (Fig. 3a), for example, by a mechanical force exerted by an atomic force microscope tip[38]. At open edges, we find reconstructions of the skyrmions, where the domain walls tend to align perpendicular to the edges (Fig. 3a). Since the doubly degenerate ground states at zero field (Fig. 2c) have different skyrmion locations (**R₁** vs **R₂**) in the moiré, the degeneracy can be lifted by their distinct edge reconstructions, as Fig. 3b shows. A topological phase transition between the skyrmion lattices of $C = 1$ and $C = -1$ therefore occurs as a function of the interlayer translation.

The moiré magnification effect also amplifies strains[39], which can be used to dramatically tune the periodicity and shape of the magnetization texture. For example, when a relative biaxial strain tunes the lattice mismatch $\eta$ between the two layers, the periodicity of moiré skyrmion lattices scales as $A = \frac{1}{\eta}a$. At small $\eta$, a tiny strain can significantly change $A$. The magnitude of the emergent electromagnetic field $\Omega$ scales inversely with $A^2$, thus the strain tuning in $A$ can manifest as a remarkable control on the Hall angle of the topological Hall effect (Fig. 3c). The strain tunability of moiré periodicity can also be exploited for controlling phase transition between the skyrmion lattice, magnetic bubble lattice, and the FM state (Fig. 2a).

*Skyrmion excitations in the FM phase.* --- Now we turn to the parameter regime of weak interlayer magnetic coupling or small moiré period where the intralayer exchange dominates. The ground state is then a FM state without domains (c.f. Fig. 2a). We find that the moiré modulated interlayer magnetic coupling, albeit weak, tends to evolve a locally created domain of reversed magnetic configuration into a metastable skyrmion. For FM ground state of negative (positive) magnetization, the metastable skyrmion excitation can be created at the **R₁** (or **R₂**) locals with topological charge $C = 1$ (or $C = -1$). The skyrmion excitation can also be eliminated through a local reversal of the magnetization at the skyrmion center. The detail of the skyrmion generation and elimination dynamics is presented in Supplementary Note 4.

Such metastable skyrmion excitations can be moved between the periodically ordered **R₁** or **R₂** locals in the moiré by current pulses, as Fig. 4 shows. To simulate the current effect on the magnetization dynamics, a spin torque term is added to Eq. (2): $\tau = -\mathbf{m}_i \times [\mathbf{m}_i \times (\mathbf{j} \cdot \nabla)\mathbf{m}_i] - \beta \mathbf{m}_i \times (\mathbf{j} \cdot \nabla)\mathbf{m}_i$.[40] The simulation based on the coupled LLG equations is performed with the parameters $A = 42a$, $J_{\text{ex}} = 0.065I$, corresponding to the point denoted by the star symbol in the FM phase region in Fig. 2a, and $\beta = \alpha = 0.2$. The skyrmion excitation has a significant increase in energy and some shape deformation when moving between two neighboring **R₁** locals in the presence of current, and then relaxes to its metastable configuration at the new **R₁** site in sub-nanosecond timescale. The simulation demonstrates that the **R₁** (**R₂**) locals in the moiré realize an ordered array of trapping sites for skyrmion excitations of topological charge $C = 1$ $(-1)$.

These possibilities to generate, erase, and move individual metastable skyrmion excitations in periodically ordered traps promise the potential use of magnets moiré as a lateral register array. Compared to existing systems where skyrmions are exploited as information carriers[29-31,34,36], the moiré skyrmion excitations features unique addressability because of the highly ordered trap locations.

*Discussions.* --- The formation of skyrmions is favored in moiré pattern with long period, which requires small lattice mismatch between the AFM substrate and the monolayer magnet. Possible candidates for the FM monolayer are the actively explored chromium trihalides ($CrX_3$, X = Cl, Br and I)[2-9]. The AFM substrate can use the layered manganese chalcogenophosphates ($MnPX_3$, X=S, Se, Te) where the lateral AFM orders are well documented in the literature[23-25] and are of revived interest for extraction of 2D AFM in the atomically thin limit[26]. The lattice mismatches are $\delta \approx 0.9\%, 1.5\%, 1.9\%$ respectively for the monolayer/substrate combination $CrCl_3/MnPS_3$, $CrBr_3/MnPSe_3$, and $CrI_3/MnPTe_3$[37,41]. The moiré pattern can be further tuned by twisting and/or strain. Moreover, with the stacking dependent sign of interlayer exchange suggested by first principle calculations[5,20-22], the twisting formed moiré between the $CrI_3$ homo-layers is another possible platform to explore the magnetization textures and skyrmions.

Magnetic anisotropy of the FM monolayer competes with the moiré patterned interlayer magnetic coupling from the substrate in determining the magnetization textures. The results presented above use a weak perpendicular anisotropy $K = 0.014J$ that corresponds to the example of monolayer $CrBr_3$. In comparison, the $CrI_3$ monolayer features stronger perpendicular anisotropy[2,42], while easy-plane anisotropy is expected for monolayer $CrCl_3$[43]. In Supplementary Figure 6, we show the phase diagrams calculated at $K = 0.057J$ and $K = -0.016J$ which correspond to the anisotropies of $CrI_3$ and $CrCl_3$ respectively[42,43]. With the increase of perpendicular anisotropy, the formation of domain walls costs more energy, so FM phase is more favored. Nevertheless, skyrmion lattice phase is still found with its boundary shifted to the stronger interlayer coupling and larger moiré period side, while the magnetic bubble phase is absent at $K = 0.057J$. Under easy-plane anisotropy, magnetic domains with alternating signs of $m_z$ can still form at relatively large interlayer exchange.

Because of the larger width of the domain walls in such case, the intralayer exchange can dominate over the interlayer dipolar field in aligning the magnetization along the domain walls, and magnetic bubble lattices are thus favored.

We have also looked into vdW heterostructure of monolayer magnet on AFM substrate of easy-plane anisotropy (e.g. MnPSe$_3$[23]). In such case, the moiré modulated interlayer magnetic coupling can also lead to skyrmion lattice phase in FM monolayer of weak perpendicular anisotropy, where the skyrmions feature a different shape and trapping location in the moiré supercell (c.f. Supplementary Figure 12).

**References**
(1) Gong, C.; Li, L.; Li, Z.; Ji, H.; Stern, A.; Xia, Y.; Cao, T.; Bao, W.; Wang, C.; Wang, Y.; Qiu, Z. Q.; Cava, R. J.; Louie, S. G.; Xia, J.; Zhang, X. *Nature* **2017**, 546, 265.
(2) Huang, B.; Clark, G.; Navarro-Moratalla, E.; Klein, D. R.; Cheng, R.; Seyler, K. L.; Zhong, D.; Schmidgall, E.; McGuire, M. A.; Cobden, D. H.; Yao, W.; Xiao, D.; Jarillo-Herrero, P.; Xu, X. *Nature* **2017**, 546, 270.
(3) Song, T.; Cai, X.; Tu, M. W.-Y.; Zhang, X.; Huang, B.; Wilson, N. P.; Seyler, K. L.; Zhu, L.; Taniguchi, T.; Watanabe, K.; McGuire, M. A.; Cobden, D. H.; Xiao, D.; Yao, W.; Xu, X. *Science* **2018**, 360, 1214.
(4) Klein, D. R.; MacNeill, D.; Lado, J. L.; Soriano, D.; Navarro-Moratalla, E.; Watanabe, K.; Taniguchi, T.; Manni, S.; Canfield, P.; Fernández-Rossier, J.; Jarillo-Herrero, P. *Science* **2018**, 360, 1218.
(5) Wang, Z.; Gutiérrez-Lezama, I.; Ubrig, N.; Kroner, M.; Gibertini, M.; Taniguchi, T.; Watanabe, K.; Imamoğlu, A.; Giannini, E.; Morpurgo, A. F. *Nat. Commun.* **2018**, 9, 2516.
(6) Ghazaryan, D.; Greenaway, M.T.; Wang, Z.; Guarochico-Moreira, V.H.; Vera-Marun, I. J.; Yin, J.; Liao, Y.; Morozov, S. V.; Kristanovski, O.; Lichtenstein, A. I.; Katsnelson, M. I.; Withers, F.; Mishchenko, A.; Eaves, L.; Geim, A. K.; Novoselov, K. S.; Misra, A. *Nature Electronics* **2018**, 1, 344.
(7) Kim, H. H.; Yang, B.; Patel, T.; Sfigakis, F.; Li, C.; Tian, S.; Lei, H.; Tsen, A. W. *Nano Lett.* **2018**, 18, 4885.
(8) Jiang, S.; Shan, J.; Mak, K. F. *Nat. Mater.* **2018**, 17, 406.
(9) Huang, B.; Clark, G.; Klein, D. R.; MacNeill, D.; Navarro-Moratalla, E.; Seyler, K. L.; Wilson, N.; McGuire, M. A.; Cobden, D. H.; Xiao, D.; Yao, W.; Jarillo-Herrero, P.; Xu, X. *Nat. Nanotech.* **2018**, 13, 544.
(10) Wang, Z.; Zhang, T.; Ding, M.; Dong, B.; Li, Y.; Chen, M.; Li, X.; Huang, J.; Wang, H.; Zhao, X.; Li, Y.; Li, D.; Jia, C.; Sun, L.; Guo, H.; Ye, Y.; Sun, D.; Chen, Y.; Yang, T.; Zhang, J.; Ono, S.; Han, Z.; Zhang, Z. *Nat. Nanotech.* **2018**, 13, 554.
(11) Deng, Y.; Yu, Y.; Song, Y.; Zhang, J.; Wang, N. Z.; Wu, Y. Z.; Zhu, J.; Wang, J.; Chen, X. H.; Zhang, Y. **2018**, arXiv:1803.02038.

(12) Ponomarenko, L. A.; Gorbachev, R. V.; Yu, G. L.; Elias, D. C.; Jalil, R.; Patel, A. A.; Mishchenko, A.; Mayorov, A. S.; Woods, C. R.; Wallbank, J. R.; Mucha-Kruczynski, M.; Piot, B. A.; Potemski, M.; Grigorieva, I.; Novoselov, K. S.; Guinea, F.; Falko, V. I.; Geim, A. K. *Nature* **2013**, 497, 594.

(13) Dean, C. R.; Wang, L.; Maher, P.; Forsythe, C.; Ghahari, F.; Gao, Y.; Katoch, J.; Ishigami, M.; Moon, P.; Koshino, M.; Taniguchi, T.; Watanabe, K.; Shepard, K. L.; Hone, J.; Kim, P. *Nature* **2013**, 497, 598.

(14) Hunt, B.; Sanchez-Yamagishi, J. D.; Young, A. F.; Yankowitz, M.; LeRoy, B. J.; Watanabe, K.; Taniguchi, T.; Moon, P.; Koshino, M.; Jarillo-Herrero, P.; Ashoori, R. C. *Science* **2013**, 340, 1427.

(15) Gorbachev, R. V.; W. Song, J. C.; Yu, G. L.; Kretinin, A. V.; Withers, F.; Cao, Y.; Mishchenko, A.; Grigorieva, I. V.; Novoselov, K. S.; Levitov, L. S.; Geim, A. K. *Science* **2014**, 346, 448.

(16) Cao, Y.; Fatemi, V.; Demir, A.; Fang, S.; Tomarken, S. L.; Luo, J. Y.; Sanchez-Yamagishi, J. D.; Watanabe, K.; Taniguchi, T.; Kaxiras, E.; Ashoori, R. C.; JarilloHerrero, P. *Nature* **2018**, 556, 80.

(17) Cao, Y.; Fatemi, V.; Fang, S.; Watanabe, K.; Taniguchi, T.; Kaxiras, E.; Jarillo-Herrero, P. *Nature* **2018**, 556, 43.

(18) Chen, G.; Jiang, L.; Wu, S.; Lv, B.; Li, H.; Watanabe, K.; Taniguchi, T.; Shi, Z.; Zhang, Y.; Wang, F. **2018**, arXiv:1803.01985.

(19) Zhang, C.; Chuu, C.-P.; Ren, X.; Li, M.-Y.; Li, L.-J.; Jin, C.; Chou, M.-Y.; Shih, C.-K. *Sci. Adv.* **2017**, 3, e1601459.

(20) Jiang, P.; Wang, C.; Chen, D.; Zhong, Z.; Yuan, Z.; Lu, Z. Y.; Ji, W. **2018**, arXiv:1806.09274.

(21) Soriano, D.; Cardoso, C.; Fernández-Rossier, J. **2018**, arXiv:1807.00357.

(22) Sivadas, N.; Okamoto, S.; Xu, X.; Fennie, C. J.; Xiao, D. **2018**, arXiv:1808.06559.

(23) Jeevanandam, P.; Vasudevan, S. *J Phys Condens Matter* **1999**, 11, 3563.

(24) Okuda, K.; Kurosawa, K.; Saito, S.; Honda, M.; Yu, Z.; Date, M. *J. Phys. Soc. Jpn.* **1986**, 55, 4456.

(25) Joy, P. A.; Vasudevan, S. *Phys. Rev. B* **1992**, 46, 5425.

(26) Du, K.-Z.; Wang, X.-Z.; Liu, Y.; Hu, P.; Utama, M. I. B.; Gan, C. K.; Xiong, Q.; Kloc, C. *ACS Nano* **2016**, 10, 1738.

(27) Wang, X.; Du, K.; Liu, Y. Y. F.; Hu, P.; Zhang, J.; Zhang, Q.; Owen, M. H. S.; Lu, X.; Gan, C. K.; Sengupta, P.; Kloc, C.; Xiong, Q. *2D Materials* **2016**, 3, 031009.

(28) Lee, J.-U.; Lee, S.; Ryoo, J. H.; Kang, S.; Kim, T. Y.; Kim, P.; Park, C.-H.; Park, J.-G.; Cheong, H. *Nano Lett.* **2016**, 16, 7433.

(29) Nagaosa, N.; Tokura, Y. *Nat. Nanotech.* **2013**, 8, 899.

(30) Wiesendanger, R. *Nat. Rev. Mater.* **2016**, 1, 16044.

(31) Fert, A.; Reyren, N.; Cros, V. *Nat. Rev. Mater.* **2017**, 2, 17031.

(32) Mühlbauer, S.; Binz, B.; Jonietz, F.; Pfleiderer, C.; Rosch, A.; Neubauer, A.; Georgii, R.; Böni, P. *Science* **2009**, 323, 915.

(33) Yu, X. Z.; Onose, Y.; Kanazawa, N.; Park, J. H.; Han, J. H.; Matsui, Y.; Nagaosa, N.; Tokura, Y. *Nature* **2010**, 465, 901.

(34) Jiang, W.; Upadhyaya, P.; Zhang, W.; Yu, G.; Jungfleisch, M.; Fradin, F.; Pearson, J. E.; Tserkovnyak, Y.; Wang, K. L.; Heinonen, O.; te Velthuis, S. G. E.; Hoffmann, A. *Science* **2015**, 349, 283.

(35) Du, H.; Che, R.; Kong, L.; Zhao, X.; Jin, C.; Wang, C.; Yang, J.; Ning, W.; Li, R.; Jin, C.; Chen, X.; Zang, J.; Zhang, Y.; Tian, M. *Nat. Commun.* **2015**, 6, 8504.

(36) Zhou, Y.; Ezawa, M. *Nat. Commun.* **2014**, 5, 4652.

(37) Chittari, B. L.; Park, Y.; Lee, D.; Han, M.; MacDonald, A. H.; Hwang, E.; Jung, J. *Phys. Rev. B* **2016**, 94, 184428.

(38) Jiang, L.; Wang, S.; Shi, Z.; Jin, C.; Iqbal Bakti Utama, M.; Zhao, S.; Shen, Y.-R.; Gao, H. J.; Zhang, G.; Wang, F. *Nat. Nanotech.* **2018**, 13 204.

(39) Tong, Q.; Yu, H.; Zhu, Q.; Wang, Y.; Xu, X.; Yao, W. *Nat. Phys.* **2017**, 13, 356.

(40) Ralph, D. C; Stiles, M. D. *J. Magn. Magn. Mater.* **2008**, 320, 1190.

(41) Zhang, W.-B.; Qu, Q.; Zhu, P.; Lam, C.-H. *J. Mater. Chem. C* **2015**, 3, 12457.

(42) McGuire, M. A.; Dixit, H.; Cooper, V. R.; Sales, B. C. *Chem. Mater.* **2015**, 27, 612.

(43) McGuire, M. A.; Clark, G.; Santosh, K. C.; Chance, W. M.; Jellison Jr, G. E.; Cooper, V. R.; Xu, X.; Sales, B. C. *Phys. Rev. Mater.* **2017**, 1, 014001.

**Acknowledgements:** We thank Ying Ran for stimulating discussions. The work is supported by the Croucher Foundation, the Research Grants Council and University Grants Committee (17303518P, AoE/P-04/08) of Hong Kong SAR.

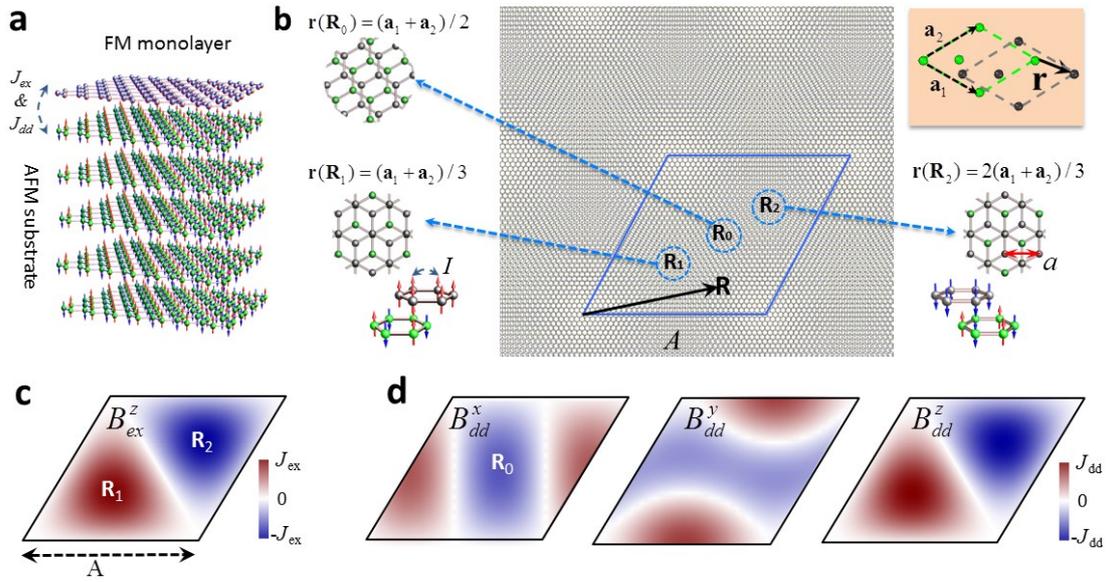

**Figure 1. Moiré modulated interlayer magnetic coupling.** (**a**) A ferromagnetic (FM) monolayer on a layered antiferromagnetic (AFM) substrate with lateral Néel order and perpendicular anisotropy. (**b**) The moiré pattern between the FM monolayer (grey) and the top layer of the AFM substrate (green) arises from the lattice mismatch and/or twisting. Three local regions at $\mathbf{R_1}$, $\mathbf{R_2}$ and $\mathbf{R_3}$ are amplified to show their local atomic registries which closely resemble the registries of lattice-matched stacking of different interlayer translations. The locally different atomic registries in the moiré can therefore be described by the interlayer translation vector **r** (c.f. inset on up right corner) as a function of location **R**. At the two locals $\mathbf{R_1}$ and $\mathbf{R_2}$, the atomic sites of FM monolayer sit on top of the two sublattices of the AFM layer with opposite magnetic order, respectively. So the interlayer exchange coupling tends to align the monolayer with opposite magnetizations at $\mathbf{R_1}$ and $\mathbf{R_2}$. (**c**) Interlayer exchange field, and (**d**) interlayer dipolar field from the AFM substrate, plotted as a function of location **R** in a moiré supercell (c.f. **b**).

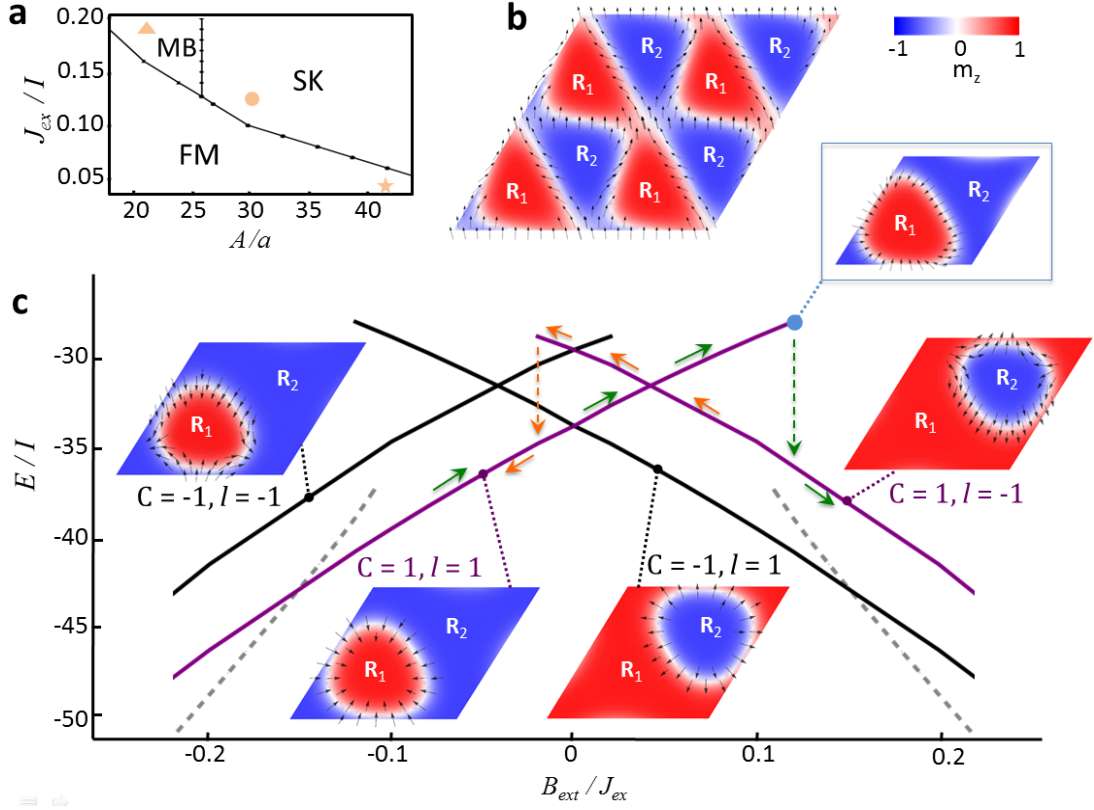

**Figure 2. Skyrmion lattice in the moiré.** (**a**) Phase diagram as a function of moiré period $A$ and the magnitude of interlayer magnetic coupling. $J_{ex}$ denotes the maximum of interlayer exchange field in the moiré. The maximum of the interlayer dipolar field z-component $J_{dd} = 0.2 J_{ex}$ (c.f. Fig. 1). The triangle in the magnetic bubble (MB) lattice phase marks parameters for part (**b**), and the dot in the skyrmion (SK) lattice phase is for part (**c**). (**b**) Magnetization texture of the MB lattice, shown in 2x2 moiré supercells. The color maps the out-of-plane component ($m_z$) of magnetization, and arrows show its in-plane orientation at domain walls. (**c**) Energies per supercell of the lowest lying stable magnetic configurations as a function of external magnetic field $B_{ext}$. Solid curves are four SK lattices of distinct topological charge C and vorticity $l$, and dashed curves are FM states. For $-0.15 \lesssim B_{ext}/J_{ex} \lesssim 0.15$, the vortex-like ($l = 1$) SK lattices are ground states, doubly degenerate at $B_{ext} = 0$. At all crossing points shown, the magnetic configurations are protected from mixing with each other by their distinct topology and vorticity. Green (orange) arrows indicate the evolution by sweeping the magnetic field up (down), where vorticity and location of skyrmion can switch under the conservation of topological charge. Inset: a vortex-like skyrmion at a critical field, which will reconfigure into an antivortex with further increase of field.

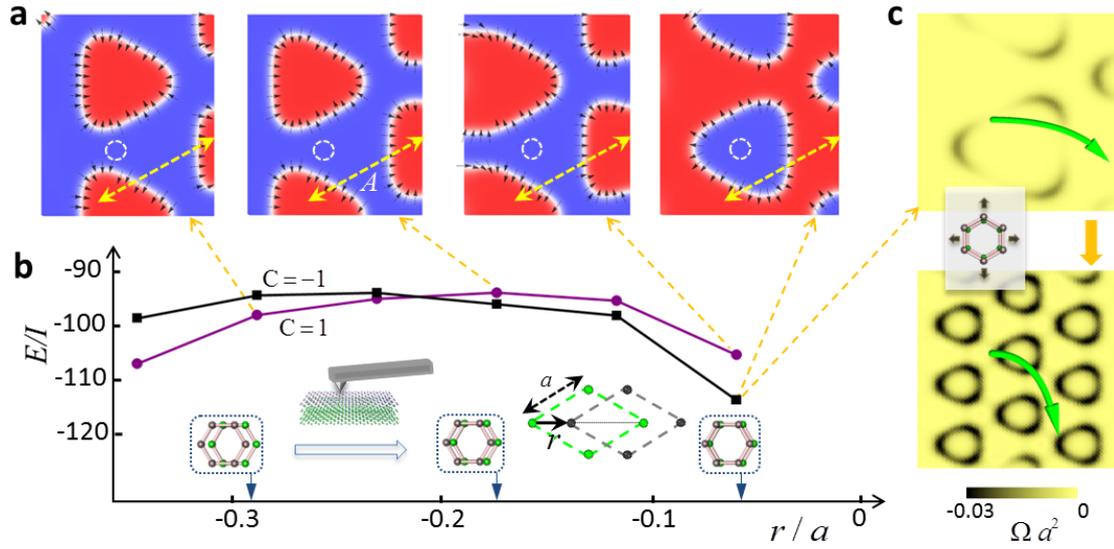

**Figure 3. Control of skyrmion lattices by interlayer translation and strain.** (**a**) Moiré magnifies a small interlayer translation $r$ into a large shift of SK lattice by $rA/a$. (**b**) The degeneracy between the ground states of opposite topological charge is lifted by the edge effect. With the sign change of the energy splitting, a topological phase transition occurs as a function of $r$, which can be controlled by a mechanical force exerted by an AFM tip. The interlayer registries atop of the $r$-axis correspond to the local atomic configuration at the dotted circles in the color map. $A/a = 67$ and open boundary condition are used. (**c**) The moiré magnetization texture leads to an emergent electromagnetic field $\Omega$ that generates a topological Hall current of carriers. A small relative strain between the two layers can lead to orders of magnitude change in the moiré periodicity, the $\Omega$ field magnitude, and hence the Hall angle.

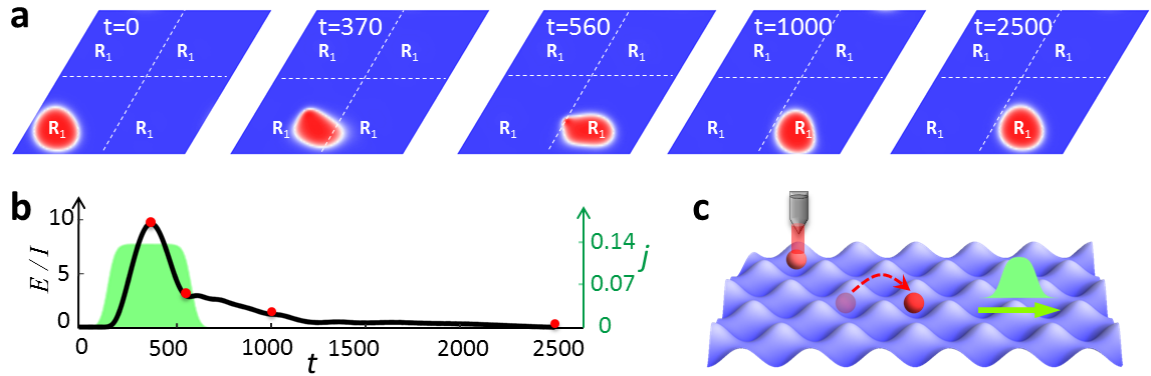

**Figure 4. Current control on the motion of skyrmion excitation.** In the FM phase, skyrmions are stable excitations that energetically favor the $R_1$ or $R_2$ locals. (**a**) Dynamics of a skyrmion excitation driven by a current pulse, shown in 2x2 supercells. (**b**) Current profile (green filled curve), and the change of skyrmion energy (black curve) as a function of time. Time is in unit of $1/(\gamma I)$, and current is in unit of $e\gamma I/(pa)$ with $p$ the spin polarization of the electric current. With the parameter values $\gamma = 1.76 \times 10^{11} \text{s}^{-1}\text{T}^{-1}$, $a = 6\text{Å}$, $p = 1$, and $I = 1.5$meV, the current pulse has a peak magnitude of $1.7 \times 10^2 A/m$ and width of ~ 100 ps. (**c**) A skyrmion behaves like a quasiparticle trapped in a superlattice potential, which can be digitally moved by current, and be written/erased using spin polarized STM.